\documentclass[aip,reprint]{revtex4-1}
\usepackage{amssymb}
\usepackage{graphicx}
\usepackage{amsmath}
\usepackage{amsfonts}
\usepackage{epsfig}
\usepackage{epstopdf}

\DeclareMathOperator{\sgn}{sgn}
\DeclareMathOperator{\Real}{Re}

\begin{document}

\title{Josephson magnetic rotary valve}

\author{I.~I.~Soloviev}
\affiliation{Skobeltsyn Institute of Nuclear Physics, Moscow State
University, Moscow, Russia} \affiliation{Lukin Scientific Research
Institute of Physical Problems, Zelenograd, Moscow, Russia}
%Skobeltsyn Lomonosov

\author{N.~V.~Klenov}
\affiliation{Physics Department, Moscow State University, Moscow,
Russia} \affiliation{Lukin Scientific Research Institute of Physical
Problems, Zelenograd, Moscow, Russia}

\author{S.~V.~Bakurskiy}
\affiliation{Physics Department, Moscow State University,
Moscow, Russia} \affiliation{Moscow Institute of Physics and
Technology, State University, Dolgoprudniy, Moscow region, Russia}
\affiliation{Faculty of Science and Technology and MESA+, Institute
for Nanotechnology, University of Twente,  Enschede, The
Netherlands}

\author{V.~V.~Bol'ginov}
\affiliation{Institute of Solid State Physics, Russian Academy of
Sciences,  Chernogolovka, Russia} \affiliation{National University of Science and Technology "MISiS", Moscow, Russia}

\author{V.~V.~Ryazanov}
\affiliation{Institute of Solid State Physics, Russian Academy of
Sciences,  Chernogolovka, Russia} \affiliation{National University of Science and Technology "MISiS", Moscow, Russia}

\author{M.~Yu.~Kupriyanov}
\affiliation{Skobeltsyn Institute of Nuclear Physics, Moscow State
University, Moscow, Russia} \affiliation{Moscow Institute of Physics
and Technology, State University, Dolgoprudniy, Moscow region,
Russia}

\author{A.~A.~Golubov}
\affiliation{Moscow Institute of Physics and Technology, State
University, Dolgoprudniy, Moscow region, Russia}
\affiliation{Faculty of Science and Technology and MESA+, Institute
for Nanotechnology, University of Twente,  Enschede, The
Netherlands}

\date{\today}

\begin{abstract}
We propose a control element for a Josephson spin valve. It is a
complex Josephson device containing ferromagnetic (F) layer in the
weak-link area consisting of two regions, representing $0$ and $\pi$
Josephson junctions, respectively. The valve's state is defined by
mutual orientations of the F-layer magnetization vector and normal
to the interface separating $0$ and $\pi$ sections of the device. We
consider possible implementation of the control element by
introduction of a thin normal metal layer in a part of the device
area. By means of theoretical simulations we study properties of the
valve's structure as well as its operation, revealing such
advantages as simplicity of control, high characteristic frequency
and good legibility of the basic states.
\end{abstract}

\maketitle

Superconducting digital circuits based on Josephson junctions
underwent significant progress in the last decades offering high
frequency data receiving and processing (e.g. all-digital RF
receiver with clock frequency of up to 30 GHz \cite{GKD}). Magnetic
flux quantization in a superconducting loop, allowing representation
of information bit as a flux quantum $\Phi_0$, is one of the key
features providing the superconducting technology advantages.
Unfortunately, a requirement to store the flux quantum in
superconducting memory cells naturally limits possibilities of their
miniaturization by geometric size of a cell $(\gtrsim30\mu^{2}),$
providing the flux quantization inside it.

There is no such restriction in magnetic devices which rely on
manipulation of local magnetizations. Their well known applications
are random access memory and recording heads \cite{F}. Recent
advances in understanding of hybrid S-F structures involving the
interplay of superconductivity (S) and ferromagnetism (F) opened
exciting opportunities for developing the new tunable Josephson
junction valves. Their operation relies on the control of induced
superconducting correlations in the weak-link area by manipulations
of magnetization of the F-layer located inside (or nearby) the
Josephson heterostructure \cite{OYB1997, Beasley2006,
BakurskiyAPL2013} or by changing the mutual orientations of
magnetization vectors of multiple F-layers \cite{Blamire2004, BFE,
Robinson1, Kohlstedt,  Baek,  Halterman, NDGWLPB}, or by making use
of the junction ground-state bistability \cite{GSWRKKK}.
Experimental realizations revealed the following drawbacks of these
approaches.

1) Proper interplay between parts of spatially inhomogeneous bistable $%
\varphi $-junction, which separately possesses the phase shifted (0 and $\pi$%
) ground states, practically limits the junction dimension to the
Josephson penetration length $\lambda _{J}$ which is larger by an
order of magnitude than the characteristic size of currently
available junctions.

2) Modulation of the effective exchange energy $H$ by mutual
misorientation of magnetization vectors $M_{1}$, $M_{2}$ of the F
films inside a SF$_{1}$F$_{2}$S structure requires combination of
strong and weak ferromagnetics in order to provide the reversal of
magnetization $M_{1}$ in just one film (with smaller coercivity)
while keeping $M_{2}$ unchanged. In this case the magnetization
reversal has little influence on magnitude of $H$ in the structure
and the critical current $I_{c}$ of the valve is highly suppressed
by $H$. This in turn leads to decrease of the $I_{c}R_{n}$ product
of the spin valve down to few microvolts\cite{Baek} ($R_{n}$ is the
valve's normal state resistance) that is to the value of three
orders of magnitude smaller than that of junctions operating in
rapid single flux quantum (RSFQ) circuits.

3) Suppression of the $I_{c}R_{n}$ product in the
SF$_{1}$F$_{2}$F$_{3}$S valves based on spin-polarized triplet
superconductivity\cite{BFE, NDGWLPB} is further enhanced (down to
nanovolt level) due to an unavoidable decrease in $I_{c}$ caused by
the increase in both: the number of non-superconducting layers and
interfaces in the weak link region.

Small magnitude of $I_{c}$ limits utilization of SF$_{1}$...F$_{n}$S
junctions as a control sFS unit of SIsFS
devices\cite{BakurskiyAPL2013} (I and s stand for an insulator and a
thin superconducting layer) which have $I_{c}R_{n}$ of millivolt
scale that is typical for RSFQ circuits.

\begin{figure}[t]
\begin{center}
\includegraphics[width=6.5cm]{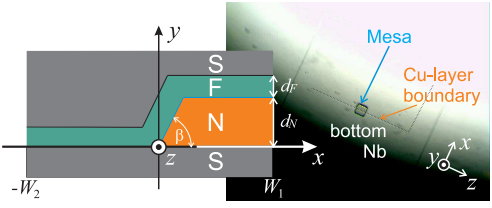}
\end{center}
\caption{Top view of the sample after CuNi-Nb bilayer deposition and
photolithography of mesa. Inset shows the sketch of cross section of
the prepared sample along the $x$ direction. Typical magnitude of
the angle $\beta$ is in the range $30^{\circ}$-$45^{\circ}$.}
\label{Fig1}
\end{figure}

To resolve the problems we restrict ourselves by a single F-layer in
the weak-link area and propose to create heterogeneity in the
contact plane providing the separation of the structure into two
regions which have positive (0 segment) and negative ($\pi$ segment)
critical current densities. If the size of the structure $L$ across
the two segments ($x$ axis in Fig.~\ref{Fig1}) is much less than
$\lambda _{J}$ then ground states of the Josephson phase in the
segments become leveled\cite{BK, GSWRKKK} $\varphi_0 = 0$ or $\pi$
(or $\pi/2$ if the critical currents of the segments are very
close). Since this ground state is $\pi$ (or $\pi/2$) shifted from
initial ground state of one (or both) of the segments, it
corresponds to nearly unstable state of the structure \cite{BK,
GSWRKKK} that manifests itself via significant reduction of the
total structure critical current $I_{c}$. We will show below that
this leveling exists only when the F-layer magnetization vector $M$
is oriented in the $x$-$y$ plane. However, if $M$ is aligned along
the boundary between the segments (along $z$ axis), the Josephson
phase gradient induced by the F-layer magnetic field shifts the values of 
$\varphi_0$ in the segments closer to their initial values, thus
removing the instability. Therefore misalignment of the F-layer
magnetization from $z$ axis up to an angle of $90^{\circ}$ should
lead to modulation of $I_{c}$ in a wide range.

Previously, the required inhomogeneity was experimentally realized
by fabricating a step in the F layer thickness in the weak link
region of SFS sandwiches \cite{SFSstep} or SIFS tunnel junctions
\cite{SIFSstep1}$^{,}$ \cite{SIFSstep4}. The devices were modeled
theoretically \cite{SIFSstepT} by considering the SIFS structures
with a constant thickness of the F film and step-like variation of
transparency of the SF interface. Both of these solutions provide a
strong coupling\cite{SIFSstep4} between the 0, $\pi $ segments in
the devices.

In this work we prove the validity of the above statement
about wide-range modulation of $I_{c}$
by studying the example of a structure in which 0 and $\pi $ segments
are formed by applying the normal film in the weak link area, as
shown in Fig.~\ref{Fig1}. The existence of an NF interface inside
the weak link region provides better decoupling of its 0 and $\pi $
segments, the property that is necessary to provide the effective
operation of the device. At the same time, the presence of the  N
film is able to provide the required $\pi$ shift in the ground
states of the segments\cite{HPKGKKRWK}. We study modulation of the
S-F/NF-S junction total critical current accompanied by changing of
its current-phase relation. Finally, we briefly discuss the
advantages of the proposed valve and a possible way of its
utilization.

The considered multilayered S-F/NF-S structure consists of
superconducting electrodes separated by either F-layer with a
thickness $d_{F}$ or by the sandwich containing combination of the
same F-layer and the N-layer having thickness $d_{N}$. To describe a
supercurrent transport we assumed that the effective electron-phonon
coupling constant is zero and that the conditions of the dirty limit
are fulfilled for both N and F metals. We also assumed that the
superconducting films are made from the same material and that the
temperature $T$ is close to the critical temperature $T_{c}$ that
permits the using of the linearized Usadel equations: \cite{Usadel}
\begin{eqnarray}
\xi _{N}^{2}\nabla^2 \mathfrak{F}_{N}-\Omega \mathfrak{F}_{N} &=&0,
\label{UsEq1} \\
\xi _{F}^{2}\nabla^2 \mathfrak{F}_{F}-\widetilde{\Omega }\mathfrak{F}%
_{F} &=&0.
\end{eqnarray}
Here $\nabla = (\partial/\partial x, \partial/\partial y)$ is the
differential operator, $\Omega =|\omega |/\pi
T_{c},\widetilde{\Omega }=\Omega +ih\sgn(\omega ) $,
$h=\mathcal{H}/\pi T_{c}$, $\omega =\pi T(2n+1)$ are the Matsubara
frequencies, $\mathcal{H} $ is the exchange energy of the F
material, $\xi _{N,F}=(D_{N,F}/2\pi T_{c})^{1/2}$ are the decay
lengths ($D_{N,F}$ are the diffusion coefficients) in N, F metals,
while $\mathfrak{F}_{N,F}$ are the Usadel Green's functions in the N
and F layers, respectively. Applying the Kupriyanov-Lukichev
boundary conditions \cite{KL} at all interfaces, we supposed that
the suppression parameter $\gamma
_{BF}=\mathcal{R}_{BF}\mathcal{A}_{BF}/ \rho _{F}\xi _{F}$ at the SF
interface is large enough to neglect suppression of
superconductivity in the S electrodes. Here $\mathcal{R}_{BF}$ and
$\mathcal{A}_{BF}$ are the resistance and the area of the SF
interfaces, $\rho _{F}$ is the resistivity of F material. Contrary
to this, the SN and the NF interfaces are supposed to be transparent
for electrons. To simplify the problem further we suppose that the
thickness of the N layer is small $d_{n}\ll \xi _{N}$, the step in
the F film is vertical \cite{suppl}, and neglect the impact of the
boundary region $ -d_{F}<x<0$ to the supercurrent. Under these
suggestions the problem can be reduce to the one-dimensional,
\cite{suppl} resulting in the superconducting current densities $J_{c}$ in SFS ($-W_{2}<x<0$%
) and SNFS ($0<x<W_{1}$) segments of the structure in the form:
\begin{subequations}
\label{UsSol}
\begin{eqnarray}
J_{c}=\frac{\mathbb{C}}{\gamma _{BF}}\Real\sum_{\omega =0}^{\infty }\frac{1}{%
\sqrt{\widetilde{\Omega }}\omega ^{2}\sinh \left( \sqrt{\widetilde{\Omega }}%
d_{F}/\xi _{F}\right) }, \label{JSF} \\
J_{c}=\Real\sum_{\omega =0}^{\infty }\sum_{m=0}^{\infty }\frac{2\mathbb{C}%
\sin \frac{k_{m}x}{\xi _{N}}}{Q\cosh \frac{\sqrt{\left\vert \omega
\right\vert +k_{m}^{2}}d_{N}}{\xi _{N}}\cosh \frac{\sqrt{\widetilde{\Omega }}%
d_{F}}{\xi _{F}}},  \label{JSNF}
&&\setcounter{MaxMatrixCols}{20}%
\end{eqnarray}
\begin{equation}
\begin{matrix}
k_{m}=\frac{\pi \xi _{N}(2m+1)}{2W_{1}}, & Q=\pi \omega ^{2}(2m+1), &
\mathbb{C}=\frac{4\pi T\Delta ^{2}}{e\gamma _{BF}\xi _{F}\rho _{F}}.\nonumber%
\end{matrix}%
\end{equation}
Here $\Delta $ is the magnitude of order parameter in the S
electrodes, $W_{1,2}$ are the lengths of the SFS and the SNFS
regions shown in Fig.~\ref{Fig1}, $e$ is the electron charge.

Figure~\ref{Fig2} gives $J_{c}(d_{F})$ dependencies of SFS and SNFS
segments of the structure. The calculations have been done for
$T=4.2$~K and for a typical experimental set of parameters
\cite{Bol}, namely
$d_{N}=20$~nm, $T_{c}=10$~K, $\mathcal{H}=32~\pi T_{c}, $ $\gamma _{BF}=0.6,$\ \ $\Delta =1.67~T_{c}$%
, %$\xi _{S}=10$~nm,
$\xi _{N}=100 $~nm, $\xi _{F}=5$~nm. It is seen that the N-layer shifts the
0-$\pi $ transitions toward the larger values of the F-layer thickness. As a
result, there are intervals of $d_{F}$ in which the critical current
densities of the parts are of opposite sign. The distinction in $%
J_{c}(d_{F}) $ curves demonstrated in Fig.~\ref{Fig2} is not due to
the finite thickness of the normal film\cite{HPKGKKRWK} but follows
from the difference in the boundary conditions at the SF and NF
interfaces.
\begin{figure}[t]
\begin{center}
\includegraphics[width=7cm]{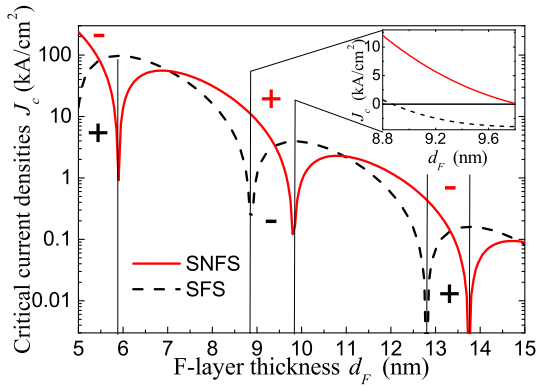}
\end{center}
\caption{Critical current densities of SFS and SNFS segments of the
structure shown in Fig.~\protect\ref{Fig1} calculated  numerically
using the expressions (\protect\ref{UsSol}). Inset presents the
densities in the range of the F-layer thickness: $d_{F} = 8.8 -
9.8$~nm.} \label{Fig2}
\end{figure}

Taking into account the step-like dependence \cite{suppl} (\ref{JSF}), (\ref%
{JSNF}) of the total critical current density $j_{c}(x)$ versus coordinate $%
x,$ we have used the two-dimensional sine-Gordon equation
\end{subequations}
\begin{equation}
{\varphi }_{tt}-{\varphi }_{xx}-{\varphi }_{zz}+j_{c}(x)\sin (\varphi
)=-\alpha {\varphi }_{t}+j-\eta _{x}-\eta _{z}  \label{SG}
\end{equation}
for simulation of the structure total critical current modulation
with rotation of the F-layer magnetization in the contact ($x-z$)
plane. The space coordinates $x$ and $z$ are normalized to $\lambda
_{J}$, the time $t$ is normalized to the inverse plasma frequency
$\omega _{p}^{-1}$, $\alpha ={\omega _{p}}/{\omega _{c}}$ is the
damping coefficient, $\omega _{p}=\sqrt{2\pi I_{c}/C\Phi _{0}}$,
$\omega _{c}=2\pi
I_{c}R_{n}/\Phi _{0}$, $C$ is the capacitance. The critical current density $%
j_{c}(x)$ as well as the overlap bias current density $j$ is
normalized to the critical current density of the SNFS segment
$J_{c\_SNFS}$. The ratio of critical current densities of the
structure segments was taken $j_{c\pi
/0}=J_{c\_SFS}/J_{c\_SNFS}=0.66$ in accordance with our theoretical
estimations (for $d_F = 9.3$~nm, see Fig.~\ref{Fig2}). Magnetization
of the F-layer $M$ enters into consideration through the components
$\eta _{x}$, $\eta _{y}$ with the standard normalization: $\eta
=2\pi\mu _{0}M\Lambda \lambda _{J}/\Phi _{0}$, where $\mu _{0}$ is
the vacuum permeability and $\Lambda $ is the
magnetic thickness of the structure.  The structure was assumed to be square shaped with the side length $%
L $ and equal lengths of its 0, $\pi $ segments $W_{1}=W_{2}$.
\begin{figure}[t]
\resizebox{1\columnwidth}{!}{
\includegraphics{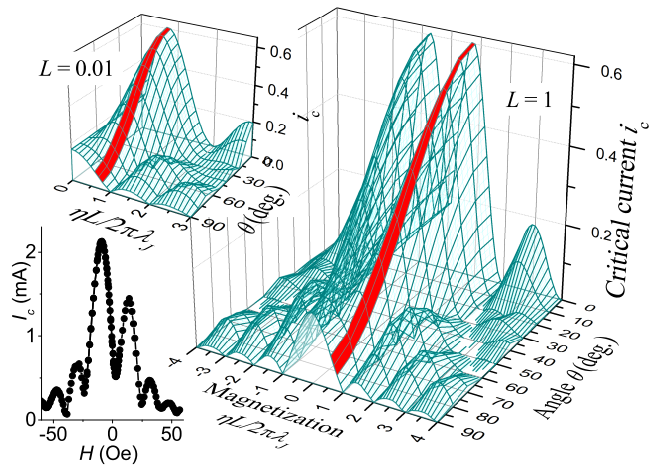}}
\caption{Total normalized critical current of the structure shown in Fig.~%
\protect\ref{Fig1} versus the F-layer magnetization $\protect\eta$
and misorientation angle $\protect\theta$ calculated  for square
shaped
structure with the side length $L=\protect\lambda _{J}$ (main panel) and $%
L=0.01\protect\lambda _{J}$ (upper inset). The valve's operational
region is painted over. The bottom inset gives experimental
dependence of critical current versus applied magnetic field
$I_c(H)$ for $\theta = 0$.} \label{Fig3}
\end{figure}

The total normalized critical current $i_c$ versus the magnetization
$\eta$ for different angles $\theta$ of the magnetization's
deviation from the $z$ axis in the $x$-$z$ plane is presented in
Fig.~\ref{Fig3}. The dependence $i_c(\eta)$ has minimum at zero
magnetization for $\theta = 0^{\circ}$. At the same time, for
magnetization oriented perpendicular to the boundary between the
segments (along the $x$ axis, $\theta = 90^{\circ}$) the dependence
has Fraunhofer-like shape that is typical for 0-$\pi $ junctions
\cite{PKKKGWFLU}. It is seen that the critical current modulation
with rotation of the magnetization is most pronounced for the
magnetization
values $\eta \lesssim 1$. To reveal the interplay between the structure 0-$%
\pi$ inhomogeneity and the F-layer magnetization value and its orientation
we calculated harmonic amplitudes of the current-phase relation (CPR): $%
i_{S}(\varphi)=A_{1}\sin\varphi+A_{2}\sin2\varphi+B_{1}\cos\varphi$.

\begin{figure}[t]
\begin{center}
\includegraphics[width=7cm]{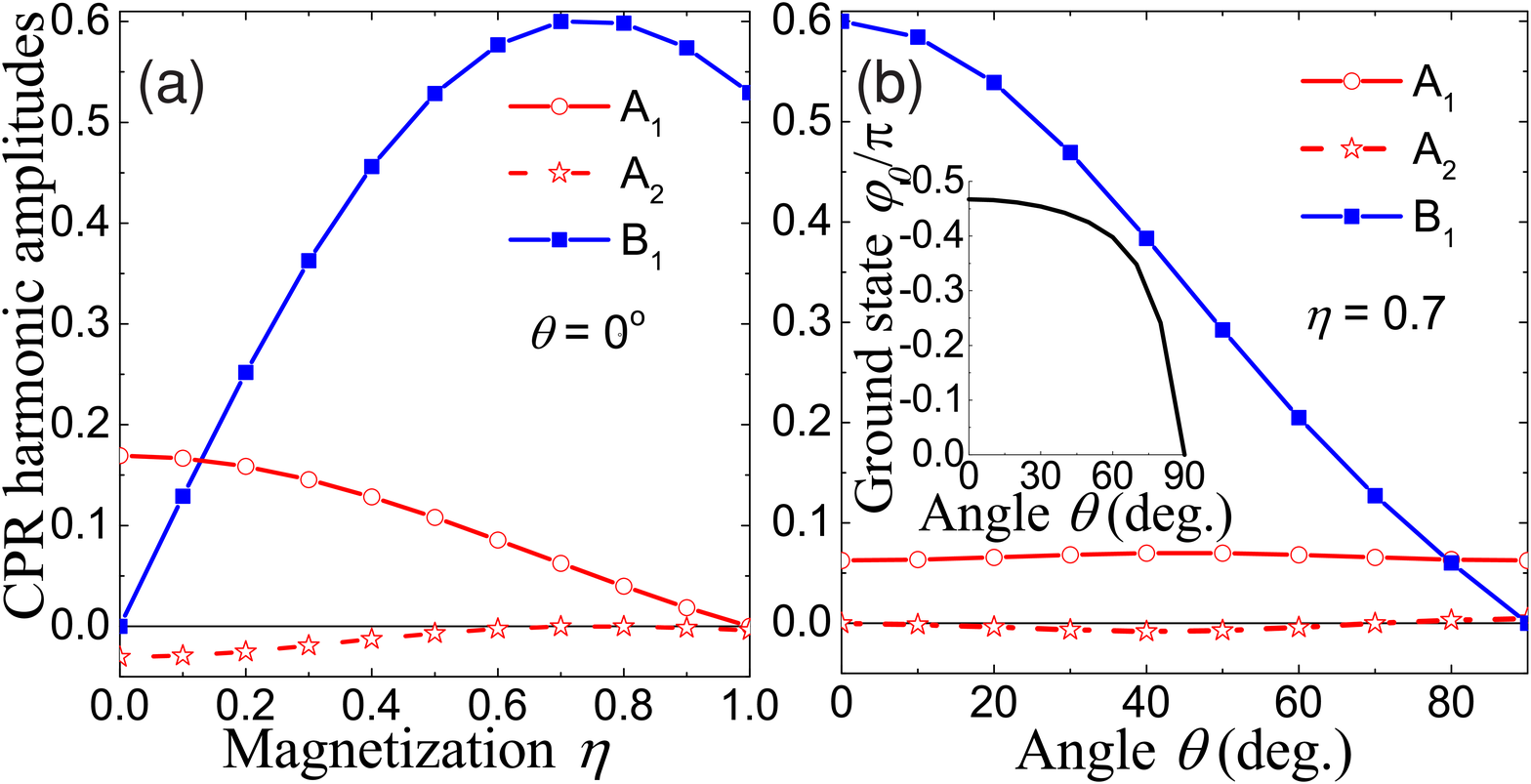}
\end{center}
\caption{(a) CPR harmonic amplitudes versus the F-layer magnetization at $%
\protect\theta=0$ and (b) the amplitudes versus the misorientation angle $%
\protect\theta$ at $\protect\eta = 0.7$. The inset presents
corresponding ground-phase shift of the structure.} \label{Fig4}
\end{figure}

Changing of the amplitudes $A_{1},$ $A_{2},$ and $B_{1}$ with the
magnetization increase $\eta =0\ldots 1$ at zero angle $\theta
=0^{\circ }$ is shown in Fig.~\ref{Fig4}a. The second harmonic in
the CPR at $\eta =0$
appears due to spontaneous modulation of the current along the junction \cite%
{BK}. In our case the amplitude of the harmonic is relatively small
$A_{2}/A_{1}=-0.18$ and insufficient for formation of a $\varphi
$-junction (this requires \cite{BK} $A_{2}<-A_{1}/2$), and therefore
the ground state is nondegenerated with $\varphi _{0}=0$. This
corresponds to the noticeable difference of the critical current
densities $j_{c\pi /0}\neq 1$ and the small dimension of the
structure \cite{PKKKGWFLU}. Increase of the magnetization induces an
asymmetry of Josephson energy potential that manifests itself in
occurrence of the cosine component in the CPR and according
ground-state shift. The total critical current increase and the
second harmonic decrease illustrate the mentioned compensation of
spontaneous phase leveling by F-layer inner magnetic field. This
effect is maximal in the vicinity of the magnetization value $\eta
=0.7$.

Modulation of the harmonic amplitudes with rotation of the
magnetization $\theta =0^{\circ }\ldots 90^{\circ }$ for $\eta =0.7$
is presented in Fig.~\ref{Fig4}b. The critical current suppression
(according curve is painted over in Fig.~\ref{Fig3}) corresponds to
the reduction of the cosine component while the sine ones are kept
nearly unchanged. This means that the $i_{c}$ modulation can be
attributed to compensation of the Josephson phase leveling across
the structure (along the $x$ axis) while the $i_{c}$ suppression due
to existence of the inner magnetic field remains the same. The valve
operation thus can be described in the frame of a simple
single-harmonic model, like the conventional RSJ \cite{RSJ} one,
but taking into account the ground-phase shift (shown in the inset of Fig.~%
\ref{Fig4}b) and the critical current modulation, which in our case is of an
order of magnitude.

We have fabricated\cite{Bol} Nb-based  S-F/NF-S junction using Cu as
the N metal and Cu$_{0.47}$Ni$_{0.53}$ as the F film. Our process
included formation of Copper rectangle on the bottom Nb followed by
ion cleaning in argon plasma and in-situ CuNi-Nb bilayer deposition.
Mesa was prepared to be splited by NF boundary (see
Fig.~\ref{Fig1}). Thicknesses of the layers and material constants
correspond to those used in the calculations. For this sample we had
observed a minimum at zero magnetic field (see the bottom inset in
Fig.~\ref{Fig3}), which is common for 0-$\pi$ junctions, while
reference SFS- and SNFS-junctions have ideal Fraunhofer-shaped
dependences of critical current versus applied magnetic field
$I_{c}(H)$. However, CuNi appeared to be an inconvenient material to
implement the valve due to its small scale domain structure,
out-of-plane magnetic anisotropy\cite{Veshchunov} and the absense of
in-plane magnetization at relatively weak magnetic fields. More
convenient ferromagnetic material is Pd$_{0.99}$Fe$_{0.01}$,
possessing opposite to CuNi properties. It
demonstrates\cite{Bolginov} sufficient magnetic hysteresis at low
magnetic fields below $10$ Oe. Experimental study of S-F/NF-S
contacts based on PdFe is an urgent task of our nearest research.

To reach the $I_{c}R_{n}$ value typical for RSFQ circuits the
considered structure can be used as a control element in SIsFS
device \cite{BakurskiyAPL2013, BKSKG}. Due to small thickness, the
intermediate s-layer can not screen the F-layer magnetic field and
therefore its rotation can control the whole SIs-F/NF-S junction
critical current. At the same time, if the s-layer thickness $d_{s}$
is much larger than its coherence length $d_{s} \gg\xi_{s}$ so that
$J_{c\_SIs} \ll J_{c\_s-F/NF-S}$ then the characteristic frequency
of the junction is determined by its SIs part and can reach the
values typical for standard SIS junctions.

In conclusion, we have suggested magnetic Josephson valve based on
interplay between spatial inhomogeneity of its structure and
orientation of its F-layer magnetization. The valve can be easily
operated by application of mutually orthogonal magnetic fields while
its states are well distinguishable due to the large critical
current modulation. The device possesses non-volatility, the ability
of non-destructive read-out, and can be used in SIs-F/NF-S structure
possessing high characteristic frequency. Since the considered
critical current modulation does not degrade with reduction of the
structure size (see upper inset of Fig.~\ref{Fig3}) the valve allows
miniaturization. Decrease of the device dimensions in the direction
normal to the boundary between its segments (along the $x$ axis)
implies proportional increase of the F-layer magnetization and
decrease of its thickness.

This work was supported by RFBR grants no. 13-02-01106, l4-02-90018-bel$\_$a, 14-02-31002-mol$%
\_$a, Ministry of Education and Science of the Russian Federation in
the frameworks of grants no. 14.616.21.0011 and 14.587.21.0006 and Increase
Competitiveness Program  of NUST «MISiS»(K2-2014-025), Russian President grant MK-1841.2014.2, Dynasty Foundation,
Scholarship of the President of the Russian Federation and Dutch FOM.

\end{document}